# Generating electromagnetic modes with fine tunable orbital angular momentum by planar topological circuits

Yuan Li[1], Yong Sun[1], Weiwei Zhu[1], Zhiwei Guo[1], Jun Jiang[1], Toshikaze Kariyado[2], Hong Chen[1*] and Xiao Hu[2★]

[1]*MOE Key Laboratory of Advanced Micro-Structured Materials, School of Physics Science and Engineering, Tongji University - Shanghai, 200092, China*

[2]*International Center for Materials Nanoarchitectonics (WPI-MANA), National Institute for Materials Science, Tsukuba 305-0044, Japan*

*email: hongchen@tongji.edu.cn

★e-mail: HU.Xiao@nims.go.jp

**Metamaterials made of periodic arrangements of electric permittivity and magnetic permeability and arrays of resonators can provide optic properties unexperienced in conventional materials, such as negative refractive index which can be used to achieve superlensing and cloaking. Developing new simpler structures with richer electromagnetic (EM) properties and wider working frequency bands is highly demanded for fast light-based information transfer and information processing. In the present work, we propose theoretically that a honeycomb-structured LC circuit with short-range textures in inductance can generate a topological photonic state accompanied by inversion between $p$- and $d$-orbital like EM modes carrying optic orbital angular momenta (OAM). We realize experimentally the topological photonic state in planar microstrips, a sandwich structure of bottom metallic substrate, middle dielectric film and top metallic strips with hexagonal patterns in strip width, in microwave frequency. Measuring accurately the amplitude and phase of EM fields, we**

**demonstrate explicitly that topological interfacial EM waves launched from a linearly polarized dipole source propagate in opposite directions according to the sign of OAM, with the relative weight of $p$- and $d$-orbital like EM modes tunable precisely by sweeping the frequency in the topological band gap, which can hardly be achieved in other photonic systems and condensed-matter systems. Achieving the topological photonic state in a simple and easy fabricable structure, the present work uncovers a new direction for generating and manipulating EM modes with rich internal degrees of freedom, which can be exploited for various applications.**

To harness at will propagation of electromagnetic (EM) waves constitutes the primary goal of photonics, the modern science and technology of light, expecting to be rewarded by valuable applications ranging from imaging and sensing well below the EM wavelength to realizing IoT, self-driving automobile and advanced information processing. So far, systems with spatially varying permittivity and/or permeability and arrays of resonators were explored, and novel EM properties unavailable in conventional uniform media have been achieved, such as negative refractive index, superlensing, cloaking and slow light[1-6], etc.

Inspired by the flourishing topological physics fostered in condensed matters[7-13], robust EM propagations at the edge of photonic topological insulators immune to back-scattering from sharp corners and imperfections came into focus in the past decade. This is achieved by one-way edge EM modes in systems with broken time-reversal symmetry (TRS)[14-23], whereas by pairs of counterpropagating edge EM modes carrying on certain pseudospins in systems respecting TRS[24-34] (for a recent review see Ref. 35). Waiving application of external magnetic field with price of sacrificing absolute robustness, topological photonic systems with TRS attract increasing interests since they are more compatible with semiconductor-based electronics and optic devices. In a topological photonic crystal with $C_{6v}$ symmetry proposed recently[27], $p$- and $d$-orbital like EM modes of opposite parities with respect to spatial inversion are tuned to generate a frequency band gap, and the sign of orbital angular momentum (OAM) plays the role of an emergent pseudospin degree of freedom. While the two EM modes with pseudospin up and down are degenerate in bulk bands due to TRS, thus hard to manipulate, they are separated into two opposite directions in the topological interface EM propagations, which can be exploited for realizing novel optic functionalities. However, up to this moment only transporting paths and integrated transmissions of topological EM edge propagations have been measured, and their detailed components remain unclear in topological photonic systems explored so far, which hampers their advanced applications (the valley degree of freedom and related OAM have been revealed in a bulk photonic graphene by selectively exciting the two sublattices in terms of interfering probe beams[36]). In this work, we present the first direct experimental observation on pseudospin states of unidirectional interface modes in topological photonic metamaterials. Based on the insight detailed by analyzing a lumped element circuit with honeycomb structure, we propose that topological propagations of EM waves can be achieved in planar microstrips and demonstrate experimentally the idea by transmission-line (TL) techniques in microwave regime (see Figure 1), using a sandwich structure of bottom metallic substrate, middle dielectric film and top patterned metallic strips[37], which is in common usage in various electronic devices. When the metallic strips form a neat honeycomb pattern, linear frequency-momentum dispersions appear in the normal frequency EM modes, much similar to the Dirac cones in the electronic energy-momentum dispersions seen in graphene. Implementing a periodic hexagonal pattern of wide/narrow metallic strips opens a frequency band gap, and particularly a band inversion between $p$- and $d$-orbitals like EM modes takes place when the inter-hexagon strips are wider than the intra-hexagon ones, results in a topological EM state mimicking the quantum spin Hall effect (QSHE) in electronic systems. Taking advantage of the planar and open structure of this metamaterial, we experimentally measure the distribution of amplitude and phase of out-of-plane electric field in terms of near-field techniques in the microwave regime, and reveal explicitly that the propagation of topological interface EM modes is governed by the sign of optic OAM, and that the relative weight of $p$- and $d$-like EM modes can be tuned precisely by sweeping the source frequency across the bulk frequency band gap. The simple structure of the present topological microstrip enables easy fabrication and direct control, which is advantageous for harnessing EM transportations and establishing novel microwave antennas to generate EM emissions with rich internal degrees of freedom. Furthermore, the open two-dimensional (2D) structure adopted in the present approach leaves much room for including other elements, such as electric/mechanical resonators, superconducting Josephson tunneling junctions and SQUID, which are useful for advanced information processing.

**Topological phase transition in lumped element circuit**

As a simplified model of our approach, we begin with a lumped element circuit shown schematically in Figure 1c with *on-node* capacitors shunt to a common ground of a uniform capacitance $C$, and *link* inductors between the nearest neighboring nodes of the honeycomb structure of two-valued inductance $L_0$ and $L_1$ inside and between hexagonal unit cells, respectively. Topological LC circuits were proposed and realized previously[33,34], where cross wirings with permutations are adopted which generates nontrivial topology. In contrary, the present approach explores the crystalline symmetry of honeycomb lattice[27,38], which makes a pure two-dimensional (2D) structure possible. The voltage on a given node with respect to the common ground is described by:

$$\frac{d^2 V_i}{dt^2} = \frac{-1}{C}\sum_{j=1}^{3}\frac{1}{L_{ij}}(V_i - V_j) \tag{1}$$

Taking the hexagonal unit cell shown in Figure 1(c), the normal frequency modes are governed by the following secular equation:

$$\left(2 + \tau - \frac{\omega^2}{\omega_0^2}\right)\tilde{V} = Q\tilde{V} \tag{2}$$

$$Q = \begin{pmatrix} 0 & Q_k \\ Q_k^\dagger & 0 \end{pmatrix}, \quad Q_k = \begin{pmatrix} \tau X\bar{Y} & 1 & 1 \\ 1 & 1 & \tau\bar{X} \\ 1 & \tau Y & 1 \end{pmatrix} \tag{3}$$

with $\vec{V} = \tilde{V}\exp(i\mathbf{k}\cdot\mathbf{r} - i\omega t) \equiv [\tilde{V}_1 \tilde{V}_2 \tilde{V}_3 \tilde{V}_4 \tilde{V}_5 \tilde{V}_6]^t \exp(i\mathbf{k}\cdot\mathbf{r} - i\omega t)$ for the voltages at the six nodes (see Supplementary Information A for details), where $X = \exp(i\mathbf{k}\cdot\mathbf{a}_1)$, $Y = \exp(i\mathbf{k}\cdot\mathbf{a}_2)$, $\omega_0^2 = 1/L_0 C$, $\tau = L_0/L_1$, and bar means complex conjugating.

Figure 2 displays the variation of frequency band structure when $\tau$ is tuned. As shown in Figure 2a, there is a global frequency band gap around 1.5 GHz for $\tau < 1$, and the EM modes exhibit double degeneracy both below and above the frequency band gap. When $\tau = 1$, the frequency band gap is closed at the $\Gamma$ point and two sets of linear dispersions, known as photonic Dirac cones, appear with four-fold degeneracy at the $\Gamma$ point as displayed in Figure 2b (accidental Dirac cones were achieved before in square lattice and used for manipulating EM transportation with zero refractive index[39]). Tuning to $\tau > 1$, a global frequency gap is reopen as shown in Figure 2c. While the frequency band structures in Figure 2a and c look similar to each other, the EM modes in the two cases are different which can be characterized by the eigenvalue of $C_2$, or equivalently the parity with respect to spatial inversion[40,41]. For $\tau < 1$ (Figure 2a), the parities of eigen EM modes are given by "$+ - -$" at both $\Gamma$ and M points. In contrary, for $\tau > 1$ (Figure 2c), the parities are given by $"+++"$ at the Γ point while $"-+-"$ at the M point below the frequency band gap at 1.5 GHz. With the parity of EM modes changed at the two high-symmetry points in the Brillouin zone (BZ), the case $\tau > 1$ features a nontrivial topology. Therefore, the lumped element circuit exhibits a topological phase transition with the pristine honeycomb structure as a transition point. Looking into detailed phase windings at the Γ point as displayed in Figure 2d-g, the EM eigen modes can be assigned as $p_\pm$ and $d_\pm$ orbitals, which are inverted between Figure 2a and c in full agreement with the change in parity of EM modes. The clockwise and counterclockwise OAM serve as the two pseudospin states in EM modes in the present approach[27-30,42-44].

While for simplicity and transparence we describe the topological phase transition by an effective lumped element circuit, it is straightforward to reformulate the physical phenomena in terms of propagating electric and magnetic fields with electric permittivity and magnetic permeability and relevant parameters. Therefore, the physics revealed here applies for a broad class of planar networks of waveguides[45], such as coaxial cables and striplines.

**Observing pseudospin and $p-d$ orbital hybridization**

We then implement experimentally the topological photonic state revealed above by designing the planar microstrip as shown in Figure 1b. Because the distributed capacitances of the metallic strips with respect to the ground plate estimated following the standard procedure[37] are smaller than the lumped ones by one order of magnitude, the distributed capacitances can be piled up to the on-node one as a sufficient approximation, resulting in the lumped element circuit discussed above. The widths of metallic strips in the trivial and topological designs are chosen in the way such that the two bulk systems give a common frequency band gap, taking into account the common lumped capacitance. In order to reveal explicitly the topological EM property, we put these two microstrips side by side as displayed in Figure 1b. As shown in Figure 3a obtained by numerical calculations based on a supercell for the lumped element circuit, two frequency dispersions appear in the common bulk frequency gap due to inclusion of the interface between the two half spaces distinct topologically. It is interesting to note that these interface modes are characterized mainly by two degrees of freedom, namely pseudospin and parity as resolved experimentally below.

In order to detect these topological interface EM modes experimentally, we launch an EM wave in terms of a linearly polarized dipole source locating at the interface with a frequency within the common bulk frequency gap (see inset of Figure 1b). It is noticed that injecting an EM wave into the system without disturbing the bulk frequency band is a feature inherent from the bosonic property of photon which is not available for electrons. As displayed in Figure 3b and c for the strength distributions of out-of-plane electric field $E_z$ obtained by full-wave finite-element simulations using CST (computer simulation technology) microwave studio software based on a finite integration method in time domain and experimental measurements, respectively, the EM wave propagates only along the interface. Figure 3d shows the phase distribution of the out-of-plane electric field $E_z$ obtained by full-wave simulations, which exhibits clockwise/counterclockwise phase winding in the left/right half of the sample to the point source, as can be seen clearly with the aid of mirror symmetry respective to a mirror line perpendicular to the interface (indicated by the arrows in Figure 3d). The two insets show the zoomed-in views of the phase distributions in two typical hexagonal unit cells close to the interface, with the left/right one accommodating the clockwise/counterclockwise phase winding, which specifies the down/up pseudospin state of EM modes. This demonstrates a clear pseudospin-momentum locking in the interface EM propagation, mimicking the helical edge states in QSHE.

Now we investigate the variation of phase distribution in the topological interface EM modes when the frequency of point source is swept across the bulk frequency band gap. The interface EM modes intersecting the frequency bands below and above the band gap are contributed from $p$ and $d$ orbitals simultaneously, which can be resolved by analyzing the phase winding (see Supplementary Information C) noticing that for a $p/d$ orbital the phase winds $2\pi/4\pi$ over a hexagonal unit cell. As displayed in Figure 3e-p obtained by full-wave simulations and experimental measurements, at a frequency close to the lower band edge (e, h, k, n) the interface EM modes are mainly contributed from $p$ orbitals, and at a frequency close to the upper band edge (g, j, m, p) the interface EM modes are predominated by $d$ orbitals, whereas $p$ and $d$ orbitals contribute equally at the center frequency of the band gap (f, i, l, o). Figure 3q displays the full frequency dependence of unnormalized weights of $p_+$ and $d_+$ orbitals evaluated in terms of a Fourier analysis on the phase distribution on the hexagonal unit cell for the right hexagon in Figure 3d (same results are obtained for the left hexagon with $p_-$ and $d_-$ orbitals as assured by the mirror symmetry), with a systematic frequency shifting of 0.03 GHz between the experimental results and the full-wave simulational ones due to the tolerance of the material and structural parameters in the fabrication. Because the $p/d$ orbitals correspond to the dipolar and quadrupolar EM modes, with a linearly polarized dipole source we can generate and guide EM waves with desired sign of OAM by choosing the emitting port, and desired relative weight of dipolar and quadrupolar EM modes by tuning the frequency within the topological band gap. These properties can be exploited for designing topology-based microwave antennas and receivers.

We can also excite topological interface EM propagations in terms of a circularly polarized source. As shown in Figure 4a and c obtained by full-wave simulations, the EM wave propagates only rightward/leftward when the point source launches a pseudospin up/down EM wave. This unidirectional EM propagation is confirmed clearly by experimental measurements as displayed in Figure 4b and d where a four-antenna array with the phase of electromagnetic wave increasingly and decreasingly by $\pi/2$ clockwise between neighboring antennas induced by delayed lines is adopted[46]. The distribution of time-averaged Poynting vector $\vec{S}$ in Figure 4e and f demonstrate the energy flows along the interface for the rightward/leftward propagating interface EM modes, respectively. As can be seen in Figure 4e with zoomed-in patterns in the left inset, the Poynting vector rotates over hexagons counterclockwise in the rightward propagating EM interface wave, same as the phase winding associated with a pseudospin up state resolved in Figure 3. In hexagons above the interface (topological regime), the density of Poynting vectors is larger at the bottom edge (closer to the interface) than that on the top edge, which generates a net rightward energy flow. In hexagons below the interface (trivial regime), although the Poynting vector rotates counterclockwise same as that in the topological regime, the density of Poynting vectors is smaller at the top edge (closer to the interface) than that on the bottom edge, opposite to that in the topological regime, which therefore also contributes a net rightward energy flow. One can see similar behaviors in Figure 4f for leftward propagating EM interface wave with a pseudospin down state. This compromises the pseudospin state, or equivalently the winding phase and rotating Poynting vector over hexagons, and the unidirectional energy flow along the interface.

As seen above, the planar and open structure of the present system permits us to observe directly the pseudospin states, pseudospin-momentum locking, and furthermore the $p-d$ orbital hybridization in the interface EM modes, which constitute the essence of the topological state preserving TRS. So far, pseudospin relevance in topological edge propagations has been inferred based on experimental observations on selective edge transportations by comparing with theoretical analyses, and there are few experimental studies on parity tuning in topological interface propagations.

**Outlook**

The topological EM properties achieved in the planar circuit can not only be exploited for microwave photonics[47] and plasmonics[48], but also be extended up to the infrared frequency regime[49]. In the photonic wire laser working at the THz band[50,51], a patterned, double-sided metal waveguide is used for confining and directing the quantum-cascade laser (QCL), with the metal-semiconductor-metal structure essentially same as the microstrip transmission line addressed in the present study. In terms of a honeycomb-patterned network with typically micrometer strip-widths one can achieve a topological QCL, where the emission and propagation of THz EM waves are governed by OAM. It is also worth noticing that the pure 2D structure in the present scheme makes topological microwave-guiding compatible with various lithographically fabricated planar devices. Extension of the lumped element circuit discussed in the present work to a network including resonators of quantum features, such as quantum bits (qubit) based on SQUID structures[52], is of special interests.

**Methods**

**Preparation of the sample and full-wave simulations.** To prepare the edges of the sample, we load lumped resistors between the metallic strips and the common ground (i.e. bottom metallic substrate), which corresponds to a perfect matching boundary condition. The values of lumped resistors are selected according to the characteristic impedances $Z_0$ of microstrip lines, 115 Ω，74 Ω，97 Ω and 66 Ω for microstrip lines with widths of 1.0 mm, 2.6 mm, 1.5 mm and 3.2 mm, respectively[37]. In order to numerically simulate the system, we perform three-dimensional (3D) full wave finite-element simulations using CST (computer simulation technology) microwave studio software based on a finite integration method in time domain. The dielectric loss tangent (tan *δ*) of the substrate and the conductivity of the metallic microstrip lines are set to be 0.0079 S/m and $5.8 \times 10^7$ S/m, respectively. The internal resistance in each lumped capacitor is taken as 1 Ω. An open boundary condition is applied for the whole sample including the terminal resistors.

**Experimental set-up.** Signals generated from a vector network analyzer (Agilent PNA Network Analyzer N5222A) are transported into a port buried in the sample, which works as the source for the system. In order to launch EM waves with circular polarization, in addition to the linearly polarized one, a four-antenna array with the phase of electromagnetic wave increasingly and decreasingly by π/2 clockwise between neighboring antennas induced by delayed lines is adopted. A small homemade rod antenna of 2 mm length is employed to measure the out-of-plane electric field $E_z$ at a constant height of 2 mm from the microstrip lines. We make sure by full-wave simulations that the field distribution thus measured is almost same as those at the very surface of microstrip lines. The antenna is mounted to a 2D translational state to scan the field distribution over the whole system with a step of 2 mm. A finer step of 1mm is taken in order to measure accurately the field distribution in several typical hexagonal unit cells, which reveals the pseudospin structure. The measured data are then sent to the vector network analyzer. By analyzing the recorded field values, we obtain the distributions of both amplitudes and phase of the out-of-plane electric field $E_z$, which are used for analysis on detailed phase winding and weights of $p$ and $d$ orbitals in topological interface states.

**Acknowledgements**

H. Chen is supported by the National Key Research Program of China (No. 2016YFA0301101), the National Natural Science Foundation of China (Grant Nos.11234010 and 61621001). X. Hu is supported by Grants-in-Aid for Scientific Research No.17H02913, Japan Society of Promotion of Science.

(2017).

30. Yang, Y. T., Xu, Y. F., Xu, T., Wang, H. X., Jiang, J. H., Hu, X. & Hang, Z. H. Visualization of unidirectional optical waveguide using topological photonic crystals made of dielectric material. Preprint at https://arXiv.org/abs/1610.07780 (2016).

31. Cheng, X. J., Jouvaud, C., Ni, X., Mousavi, S. H., Genack, A. Z. & Khanikaev, A. B. Robust reconfigurable electromagnetic pathways within a photonic topological insulator. *Nat. Mater.* **15,** 542-548 (2016).

32. Chen, W. J., Jiang, S. J., Chen, X. D., Zhu, B. C., Zhou, L., Dong, J. W. & Chan, C. T. Experimental realization of photonic topological insulator in a uniaxial metacrystal waveguide. *Nat. Commun.* **5,** 5782 (2014).

33. Albert, V. V., Glazman, L. I. & Jiang, L. Topological properties of linear circuit lattices. *Phys. Rev. Lett.* **114,** 173902 (2015).

34. Ningyuan, J., Owens, C., Sommer, A., Schuster, D. & Simon, J. Time- and site-resolved dynamics in a topological circuit. *Phys. Rev. X* **5,** 021031 (2015).

35. Khanikaev, A. B. & Shvets, G. Two-dimensional topological photonics. *Nat. Photon.* **11**, 763-773 (2017)

36. Song, D. H., Paltoglou, V., Liu, S., Zhu, Y., Gallardo, D., Tang, L. Q., Xu, J. J., Ablowitz, M., Efremidis, N. K. & Chen, Z. G. Unveiling pseudospin and angular momentum in photonic graphene. *Nat. Commun.* **6,** 6272 (2015)

37. Hong, J. S. & Lancaster, M. J. *Microstrip Filters for RF/Microwave Application* (Wiley, New York, 2001).

38. Fu, L. Topological crystalline insulators. *Phys. Rev. Lett.* **106,** 106802 (2011).

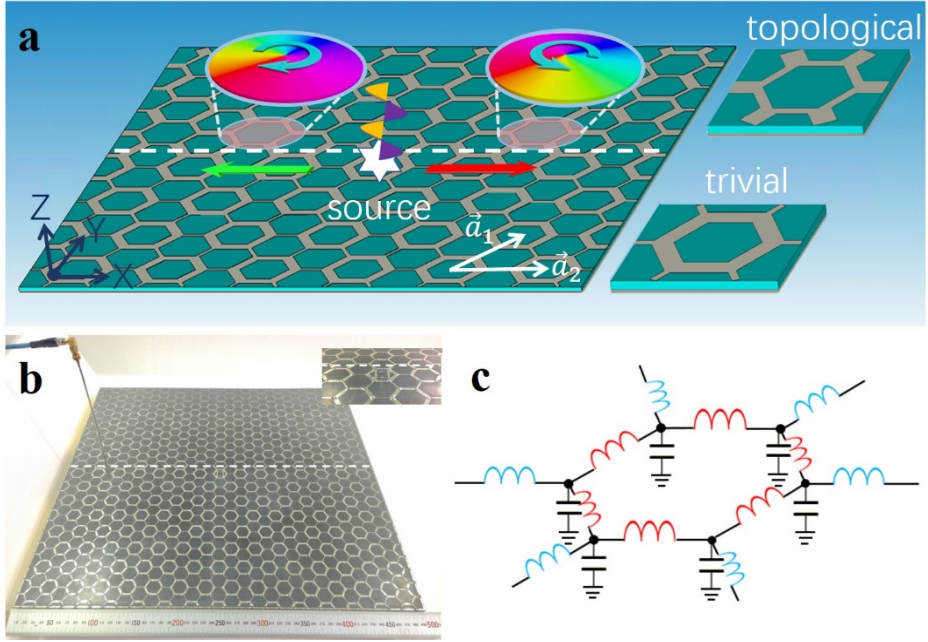

**Figure 1 | Design principle of microstrip-based topological metamaterial. a**, Schematics of the microstrip of honeycomb structure with enlarged views of the topologically nontrivial (upper) and trivial (lower) unit cells shown in the right panels. When excited by a linearly polarized source locating on the interface with frequency within the bulk frequency band gap, EM waves propagate rightward/leftward (red/green arrow) along the interface governed by the up/down pseudospin, which are represented by the phase winding of the out-of-plane electric field $E_z$ accommodated in the hexagonal unit cells as indicated in the insets. **b**, Photo for the experimental setup with a field probe placed right above the microstrip, which is used to measure the distribution of amplitude and phase of the out-of-plane electric field $E_z$, and resolve the pseudospin states and pseudospin-dominated unidirectional interface EM propagation. A lumped capacitor of $C = 5.6$ pF is loaded on the nodes. In the lower half of the system, the metallic strips of inter/intra hexagonal unit cell are of widths 1 mm and 2.6 mm, whereas in the upper half they are of 3.2 mm and 1.5 mm, respectively, and at the interface the width of metallic strips is taken as 2.6 mm. The length of all metallic strip segments is 10.9 mm and both lower and upper halves

are composed by 14x8 hexagons. The whole microstrip system is fabricated on F4B dielectric film with thickness 1.6 mm and relative permittivity 2.2. **c**, Schematic of lumped element circuit with the hexagonal unit cell shown in the right panels of **a**.

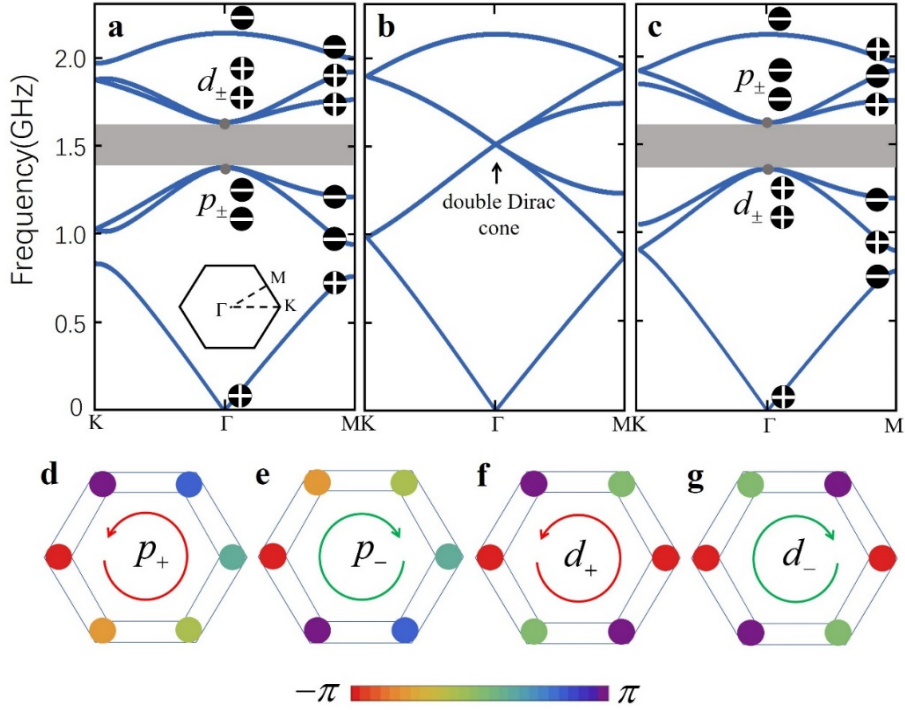

**Figure 2 | Frequency band structures and topological phase transition in lumped element circuit.** Frequency band structures calculated in terms of equation (2) and (3) for **a**, $\tau < 1$ with $L_0 = 3.60$ nH and $L_1 = 6.35$ nH, **b**, $\tau = 1$ with $L_0 = L_1 = 4.22$ nH, and **c**, $\tau > 1$ with $L_0 = 5.09$ nH and $L_1 = 3.13$ nH. The on-node capacitance is taken as $C = 7.27$ pF for all three cases. The lumped inductances and capacitance are taken as tuning parameters which reproduce the frequency band gap and gap-center frequency of the experimental setup. The values of inductances are close to those evaluated from the structures of microstrip according to standard procedures, and the on-node conductance is slightly larger than the lumped one due to the distributed capacitances from microstrip lines. The signs " + " and " − " denote the parity of the eigen EM modes with respect to spatial inversion at the high-symmetry Γ point and M point of the BZ. **d, e, f** and **g** Phase distributions of the out-of-plane electric field $E_z$ for the four eigen modes at the Γ point close to the frequency band gap at 1.5 GHz in **a** and **c**.

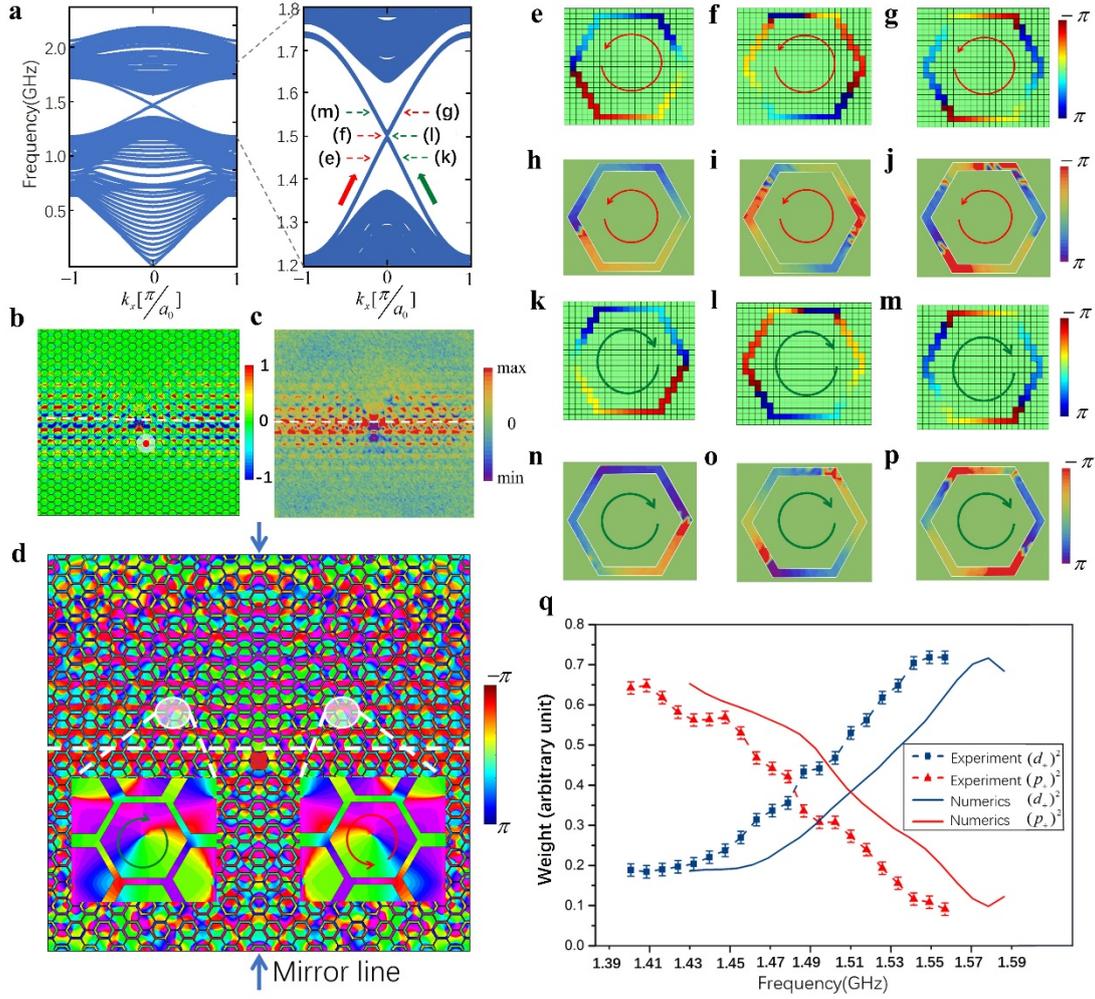

**Figure 3 | Resolving pseudospin, pseudospin-momentum locking and $p-d$ orbital hybridization. a**, Calculated frequency band structure for the whole system in Figure 1**b** with an interface between the topologically trivial and nontrivial regimes. A supercell is adopted including 8 hexagonal unit cells on both sides of the interface where the parameters for Figure 2a and c are taken respectively, presuming a system infinite along the horizontal direction. The right panel is a zoomed-in view of the frequency band diagram around the bulk band gap at 1.5 GHz. The red/green arrow indicates the dispersion of the rightward/leftward-propagating interface state. **b** and **c**, Strength distributions of the out-of-plane electric field $E_z$ obtained by full-wave simulations and experimental measurements excited by a linearly polarized source located at the interface. The source frequency is set $f = 1.47$ GHz for full-wave simulations and $f = 1.44$ GHz for

experimental measurements with a frequency shift 0.03 GHz due to the tolerances of the material and structural parameters in the fabrication. **d**, Phase distribution of the out-of-plane electric field $E_z$ under the same condition as **b**, which is mirror symmetric with respect to the mirror line perpendicular to the interface indicated by the arrows. The two insets show the zoomed-in views of the patterns of phase distribution in two typical hexagonal unit cells close to the interface, with the left/right one accommodating clockwise/counterclockwise phase winding, which indicates clearly the down/up pseudospin state associated with the leftward/rightward propagating interface EM wave. **e**, **f**, **g** and **h**, **i**, **j**, Full-wave simulated and experimentally measured phase distributions in the right highlighted hexagon in **d** with up pseudospin at three frequencies indicated in the right panel in **a**. **k**, **l**, **m** and **n**, **o**, **p**, Same as those in **e**, **f**, **g** and **h**, **i**, **j** respectively except in the left highlighted hexagon in **d**. **q**, Frequency dependence of unnormalized weights of $p$ and $d$ orbitals obtained by full-wave simulations and experimental measurements for the right hexagonal unit cell in **d**. The $p$ and $d$ orbitals take the same weight at the frequency where the two interface frequency dispersions cross each other in **a**, with an apparent difference of 0.03 GHz between the simulated and experimental results.

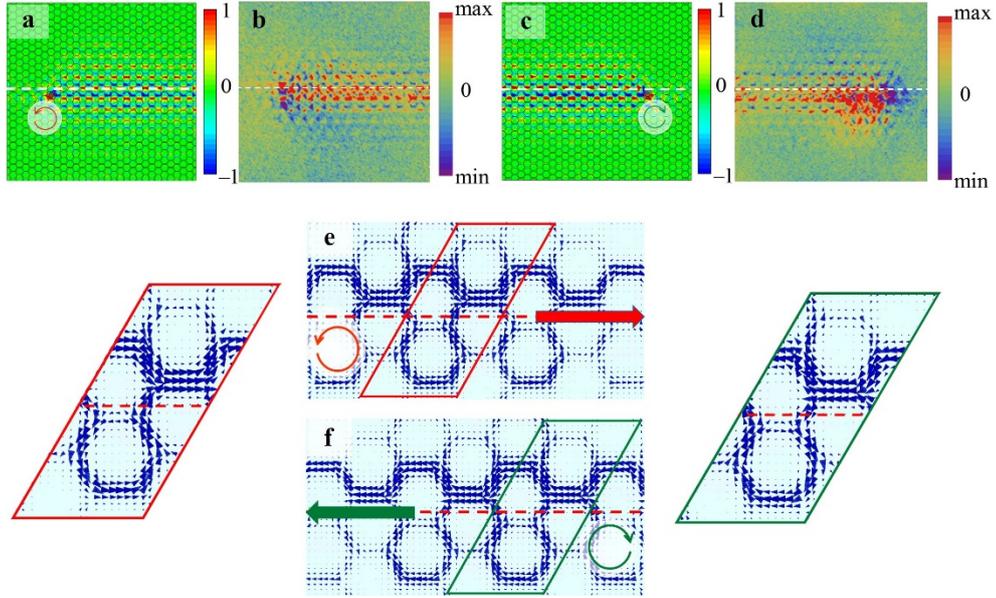

**Figure 4 | Experimental verification of interface pseudospin-unidirectional propagation.** **a** and **c**, Strength distribution of the out-of-plane electric field $E_z$ obtained by full-wave simulations excited by a point source at 1.47 GHz with circular polarization corresponding to pseudospin up and down state, respectively. **b** and **d**, Those obtained by experimental measurements at 1.44 GHz, where the four-antenna array with phase of electromagnetic wave increasing and decreasing by $\pi/2$ clockwise between neighboring antennas is used to excite the pseudospin up and down state, respectively. **e** and **f**, Distribution of time-averaged Poynting vector $\vec{S}$ of the rightward and leftward propagating interface EM wave (red and green arrow), respectively, obtained by full-wave simulations with two zoomed-in patterns.